\documentclass[letterpaper,reprint, aps,notitlepage,superscriptaddress, twocolumn,prl]{revtex4-2}
\usepackage{babel}
\usepackage[T1]{fontenc}
\usepackage[utf8]{inputenc}
\usepackage{units}
\usepackage{amsmath}
\usepackage{amssymb}
\usepackage{graphicx}
\usepackage[dvipsnames]{xcolor}
\usepackage{colortbl}

\makeatletter

\pdfpageheight\paperheight
\pdfpagewidth\paperwidth

\colorlet{mylinkcolor}{teal}
\colorlet{mycitecolor}{teal}
\colorlet{myurlcolor}{teal}

\usepackage{hyperref}
\hypersetup{
	linkcolor  = mylinkcolor,
	citecolor  = mycitecolor,
	urlcolor   = myurlcolor,
	colorlinks = true,
	breaklinks = true
}

\usepackage{tikz}
\usetikzlibrary{calc}

\usepackage{scalerel}
\usepackage{tikz}

\def\sasin{PuDTAI}
\def\swft{QMTI}
\def\methods{Methods}

\def\hgz{\scalerel*{\begin{tikzpicture}
		\fill [red, domain=-0.5:0.5, variable=\x]
		(-0.5, 0)
		-- plot ({\x}, {exp(-30*\x*\x)})
		-- (0.5, 0)
		-- cycle;
		\draw[scale=1, domain=-0.5:0.5, smooth, variable=\x, black, line width=0.5mm] plot ({\x}, {exp(-30*\x*\x)});
		\end{tikzpicture}}{p}}
\def\hgo{\scalerel*{\begin{tikzpicture}
		\fill [blue, domain=-0.5:0, variable=\x]
		(-0.5, 0)
		-- plot ({\x}, {\x*exp(-30*\x*\x)*sqrt(30)})
		-- (0.5, 0)
		-- cycle;
		\fill [red, domain=0:0.5, variable=\x]
		(0.5, 0)
		-- plot ({\x}, {\x*exp(-30*\x*\x)*sqrt(30)})
		-- (0.5, 0)
		-- cycle;
		\draw[scale=1, domain=-0.5:0.5, smooth, variable=\x, black, line width=0.5mm] plot ({\x}, {\x*exp(-30*\x*\x)*sqrt(30)});
		\end{tikzpicture}}{p}}

\makeatother

\begin{document}
\title{Optical-domain spectral super-resolution via a quantum-memory-based
time-frequency processor}
\author{Mateusz Mazelanik}
\email{m.mazelanik@cent.uw.edu.pl}

\affiliation{Centre for Quantum Optical Technologies, Centre of New Technologies,
University of Warsaw, Banacha 2c, 02-097 Warsaw, Poland}
\affiliation{Faculty of Physics, University of Warsaw, Pasteura 5, 02-093 Warsaw,
Poland}
\author{Adam Leszczyński}
\affiliation{Centre for Quantum Optical Technologies, Centre of New Technologies,
University of Warsaw, Banacha 2c, 02-097 Warsaw, Poland}
\affiliation{Faculty of Physics, University of Warsaw, Pasteura 5, 02-093 Warsaw,
Poland}
\author{Michał Parniak}
\email{m.parniak@cent.uw.edu.pl}

\affiliation{Centre for Quantum Optical Technologies, Centre of New Technologies,
University of Warsaw, Banacha 2c, 02-097 Warsaw, Poland}
\affiliation{Niels Bohr Institute, Universtiy of Copenhagen, Blegdamsvej 17, 2100
Copenhagen, Denmark}
\begin{abstract}
Existing super-resolution methods of optical imaging hold a solid
place as an application in natural sciences, but many new developments
allow for beating the diffraction limit in a more subtle way.
One of the recently explored strategies to fully exploit information already
present in the field is to perform a quantum-inspired tailored measurements.
Here we exploit the full spectral information of the optical
field in order to beat the Rayleigh limit in spectroscopy. We employ
an optical quantum memory with spin-wave storage and an embedded processing
capability to implement a time–inversion interferometer for input
light, projecting the optical field in the symmetric–antisymmetric
mode basis. Our tailored measurement achieves a resolution of 15 kHz
and requires 20 times less photons than a corresponding Rayleigh-limited
conventional method. We demonstrate the advantage of our technique
over both conventional spectroscopy and heterodyne measurements, showing
potential for application in distinguishing ultra-narrowband emitters,
optical communication channels, or signals transduced from lower-frequency
domains. 
\end{abstract}
\maketitle
Optical spectroscopy is an indispensable tool in the studies of matter
and light, including chemistry \citep{Orrit2014,Bec2020}, astronomy
\citep{Kitchin1995}, biology and medicine \citep{Kim2020}, metrology,
and more. Yet, the resolution of all state-of-the-art methods, such
as grating-based \citep{Cheng2019,Savage2009} and Fourier spectrometers
\citep{Hashimoto2018,Watanabe2018,Paudel2020}, is subject to the
Fourier limit.  Methods of beating analogous Rayleigh limit are widely known in
the context of imaging and include modifying, or exploiting very specific
properties of the source or illumination \citep{Dertinger2009,Rust2006,Betzig2006,GattoMonticone2014},
which is often impossible to implement. Furthermore, even though the Rayleigh limit
has been originally formulated in the context of a spectroscope \citep{F.R.S1879},
super-resolution methods of spectroscopy are scarce and limited to
laser spectroscopy \citep{Boschetti2020}. The task and challenges
of fluorescence spectroscopy are starkly different: a typical emitter
provides only a small photon number per single spectro-spatial mode,
which is a strong incentive to seek quantum-enhanced protocols.

\begin{figure*}[t]
	\centering{}\includegraphics[width=1.7\columnwidth]{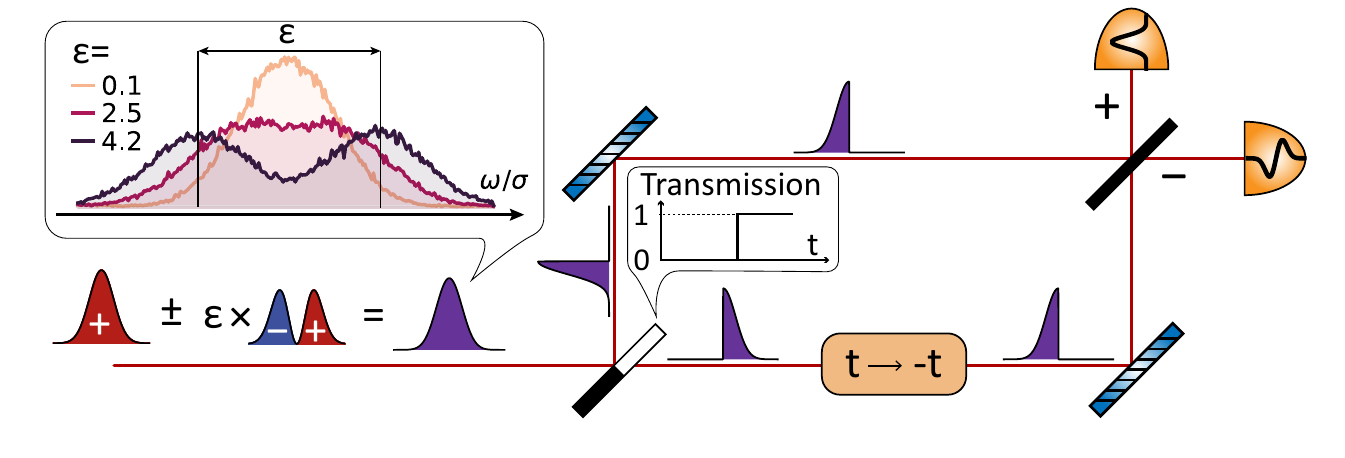}\caption{\textbf{The idea for time-inverting interferometer for superresolving spectroscopy.} The task
		is to measure a separation between two spectral lines. Separation
		is smaller than the linewidths, which are determined by the properties
		of the source itself or by the temporal aperture of the spectroscopic measurement
		device. As shown in the inset, photon shot noise (stochastic counts) prevent accurate estimate of separation when to two lines overlap strongly. For small separations, however, the information about the separation is all contained in the antisymmetric part of the signal pulse. To extract this information, the signal pulse is sent to the pulse-division time-axis-inversion
		(PuDTAI) interferometer that performs decomposition onto symmetric and antisymmetric
		components. First, a time-dependent-transmission mirror splits the pulse
		in half between the two interferometer arms. One of the arms includes
		a time inversion device that mirrors the first half of the pulse.
		Finally, the two components are combined on a beamsplitter, and the
		resulting symmetric ($+$) and antisymmetric ($-$) projections are detected.}\label{fig:pudtai}
\end{figure*}

In imaging, the modern understanding of the Rayleigh limit is formulated
as a vanishing information about separation distance between two sources for small
separations, termed the ``Rayleigh curse'' \citep{Tsang2016}. Tsang et al.
\citep{Tsang2016} have noticed that the quantum Cramér-Rao bound
(Q-CRB), which identifies maximum information available in the optical
field, is not saturated by traditional methods. This work inspired
new methods that allow for reaching the quantum limit \citep{Tsang2017,Yang2017,Lupo2020,Tham2017,Nair2016,Nair2016a,Paur2016},
most recently also in the time domain \citep{Donohue2018,Ansari2021}.
The idea to apply an analogous protocol for spectroscopy faces challenges,
especially if operation with narrowband optical signal is desired.
Yet, the inspiration can be drawn from another domain: in the nuclear-magnetic
resonance (NMR) spectroscopy a quantum memory can lead to increased
resolution when sensing with color centers in diamond \citep{Zaiser2016,Rosskopf2017,Glenn2018}.
A general framework for those experiments has been formulated by Gefen
et al. \citep{Gefen2019} who identified specific quantum measurements
that facilitate super-resolution. A resulting idea is therefore to
use a quantum memory to achieve spectral super-resolution in the optical
domain.

Here we bring the quantum-inspired super-resolution methods to the
spectral domain and demonstrate a device that can resolve frequency
differences of two emitters with precision below the Fourier limit. Here, by analogy to modern works on imaging, we understand this limit as a vanishing information about frequency separation between two spectral lines.
Our method utilizes a Gradient Echo Memory (GEM) with built-in time-frequency
processing capabilities which we program to realize a pulse-division
time-axis-inversion (\sasin) interferometer. Our protocol operates
in the ultra-narrowband domain, achieving a resolution of 15 kHz with
simultaneous super-resolution enhancement factor of $\mathfrak{s}=20\pm 0.5$
which means that about 20 times less photons are needed to achieve the
same resolution as direct spectroscopy under the same experimental conditions.  
Our work not only establishes a new super-resolution spectroscopy
method, but also provides an high spectral resolution
in absolute terms. Fundamentally, this super-resolving method exploits
the spectral information already present in light, not requiring specific
properties of source or illumination. 
Such performance is achieved by employing a spin-wave quantum memory to fully extract both phase and amplitude from the optical field. This is enabled by the the long coherence time of the memory that allow us to capture, process and release the light thus allowing optimal detection.

\section*{Results}
\subsection*{Information in spectral resolution}

Let us first set the stage of the problem and introduce the theoretical framework to evaluate the super-resolution
enhancement. Two mutually incoherent sources with equal brightness emit
photons with the same spectral envelope $\tilde{\psi}(\omega)$ that
is assumed to be known, and is the Fourier transform of the time-domain
envelope $\psi(t)$.  Such scenario is relevant both for the cases
of fluorescence (with photon pulses being determined by spontaneous
emission, assuming fast excitation and no inhomogenous broadening)
and scattering (such as Raman scattering, where we may need to consider
relative phases of ground-state coherences). It is also an optical
analogue of the radio-frequency-domain physics present in spectrum
analyzers and nano-NMR \citep{Gefen2019}. A more conventional, but equally relevant scenario is when the linewidths
of two sources are very narrow, and the resolution is limited by
an aperature, which sets the transfer function $\tilde{\psi}(\omega)$.

The two sources in question have slightly different central frequencies
$\omega_{-}=\omega_{0}-\delta\omega/2$ and $\omega_{+}=\omega_{0}+\delta\omega/2$,
where the frequency separation $\delta\omega=\sigma\varepsilon$ is
smaller than the width $\sigma$ of the spectral envelope $\tilde{\psi}(\omega)$
that for the normalized separation parameter $\varepsilon$ translates
to $\varepsilon\ll1$. The spectroscopist's task is to estimate the
separation $\varepsilon$ with maximum efficacy - to obtain maximum
information about the separation per collected photon.

\begin{figure*}[t]
	\centering{}\includegraphics[width=\textwidth]{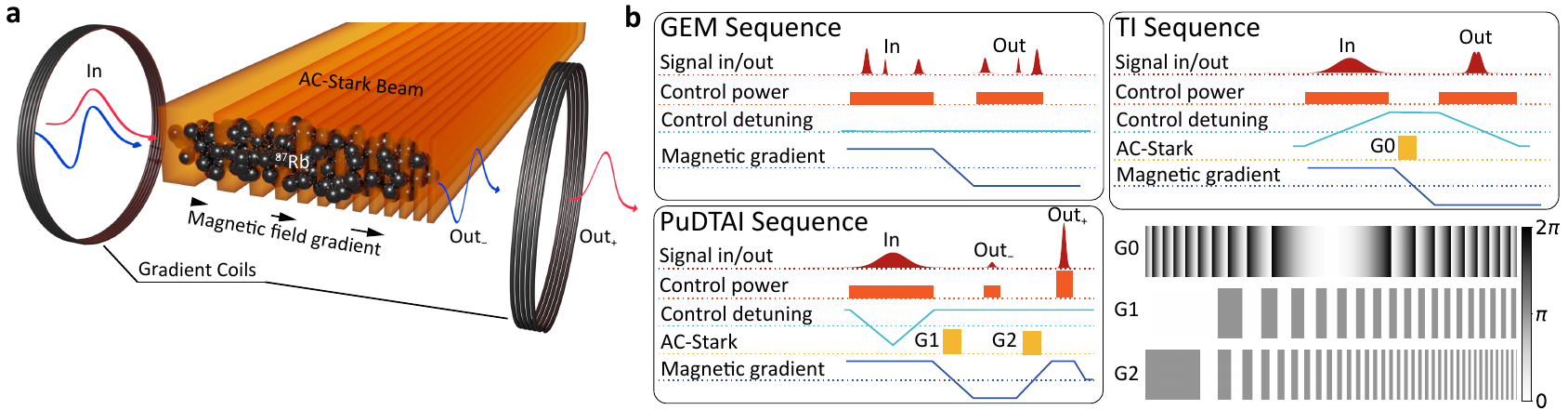}\caption{\textbf{Experimental setup and sequence.} \textbf{a} The Gradient Echo Memory (GEM) device enhanced with the spatial AC-Stark modulation, used to store, process and release the signal pulses. The same setup is used to implement the simple GEM, the temporal imaging (TI) and the \sasin\ protocol, for which the pulses are shown here. In the \sasin\ interferometer implementation a symmetrically-chirped control field maps the signal pulse onto atomic ground-state coherence of $^{87}$Rb atoms and simultaneously implements the time-dependent-transmission mirror and time-inversion transformation. AC-Stark phase modulation with chirped square wave gratings (AC-S Gratings) is used to implement the beamsplitter operation and sequentially read-out the resulting interferometer ports ($\mathrm{Out}_{-}$, $\mathrm{Out}_{+}$). \textbf{b} Experimental sequences and AC-Stark beam patterns (G0-G2) used to implement the basic (pass through) GEM protocol, TI spectrometer and the \sasin\ interferometer.  The full sequence for \sasin\ protocol and relevant atomic levels are shown in Figure \ref{fig:Experimental-timing-sequence} in the Methods section.\label{fig:Experimental-setup}}
\end{figure*}

In the conventional approach a spectrometer that measures
the normalized spectral intensity $\tilde{I}(\omega)=\frac{1}{2}\left(|\tilde{\psi}(\omega-\delta\omega/2)|^{2}+|\tilde{\psi}_{-}(\omega+\delta\omega/2)|^{2}\right)$
is used to collect data that is then processed to estimate $\delta\omega$, for example by fitting the theoretical curve. Similarly, a Fourier spectrometer
measures the second-order autocorrelation function, which is then Fourier-transformed
to yield $\tilde{I}(\omega)$. 
Notably, from the informational point of view one may obtain precision of $\delta\omega$ much greater than $\sigma$, simply by collecting enough statistics. 
Therefore, to quantify the metrological advantage, we need to consider photon shot noise resulting from finite statistics, for which the typical asymptotic behaviour is $\Delta^2 \hat{\varepsilon} \propto 1/N$ with $N$ being the number of collected photons. This behaviour is captured by the Cramér-Rao bound, which limits the maximum achievable precision for any locally unbiased
estimator of the normalized separation $\hat{\varepsilon}$ \citep{Kay1993}:
\begin{equation}
\Delta^{2}\hat{\varepsilon}\geq\frac{1}{N\mathcal{F}_{\varepsilon}},\mathcal{F}_{\varepsilon}=\int\frac{1}{p_{\varepsilon}(x)}\left(\frac{\partial}{\partial\varepsilon}p_{\varepsilon}(x)\right)^{2}\mathrm{d}x,\label{eq:CRBandFisher}
\end{equation}
where $\Delta^{2}\hat{\varepsilon}$ is the variance of the estimator,
$\mathcal{F}_{\varepsilon}$ is the Fisher information, $N$ is the
number of independent single photons used, and $p_{\varepsilon}(x)$
represents the measurement outcome distribution parametrized by the
true separation value $\varepsilon$. 
For Gaussian envelopes 
\begin{equation}
\tilde{\psi}_{\hgz}(\omega)=\left(\sqrt{2\pi}\sigma\right)^{-1/2}\exp\left(-\frac{\omega^{2}}{4\sigma^{2}}\right)\label{eq:envelopa}
\end{equation}
the Fisher information for DI in case of small separations $\varepsilon\ll1$
can be approximated as $\mathcal{F}_{\mathrm{DI}}\approx\varepsilon^{2}/8$. This unfavourable scaling has been termed the Rayleigh curse, therefore we consider DI to be a Rayleigh-limited method. Physically, this reflects the fact that it is extremely hard to estimate the separation of two noisy, overlapping spectral lines as shown in the inset in Fig. \ref{fig:pudtai}. With more statistics, one may be able to do it, but the per-photon information $\mathcal{F}_\mathrm{DI}$ is small.
At the same time the quantum Fisher information (QFI) \citep{Helstrom1976}
that yields a measurement-strategy-independent precision bound for
given estimation task  turns out to have constant value $\mathcal{F}_{\mathrm{Q}}=1/4$, independent of $\varepsilon$.
Such analysis leads to the conclusion that the direct spectroscopy scenario, 
which is an analogy for real space direct imaging (DI) of two incoherent
sources, is not optimal \citep{Tsang2016}. Hence, any strategy with Fisher information above $\mathcal{F}_{\mathrm{DI}}$ can be considered a super-resolving method beating the Rayleigh limit, or equivalently the Fourier limit in spectroscopy.

These observations have led to extensive search for different measurement
schemes that will approach the ultimate bound given by Q-CRB and
will not suffer from Rayleigh curse. Most of those super-resolution
measurement schemes are based on the following observation: a shifted
Gaussian mode function $\tilde{\psi}_{\hgz}(\omega\pm\delta\omega/2)$
for $\varepsilon\ll1$ can be approximated as an unshifted component
$\tilde{\psi}_{\hgz}(\omega)$ with a small correction of the first
Hermite-Gaussian function: 
\begin{gather}
\tilde{\psi}_{\hgz}(\omega\pm\varepsilon\sigma/2)\approx\tilde{\psi}_{\hgz}(\omega)\pm\frac{\varepsilon}{4}\tilde{\psi}_{\hgo}(\omega)\\
\tilde{\psi}_{\hgo}(\omega)=\frac{\omega}{\sigma}\left(\sqrt{2\pi}\sigma\right)^{-1/2}\exp\left(-\frac{\omega^{2}}{4\sigma^{2}}\right).
\end{gather}
 From this we see that only the antisymmetric component $\tilde{\psi}_{\hgo}(\omega)$ carries the information about the separation,
thus filtering out the (orthogonal) symmetric mode $\tilde{\psi}_{\hgz}(\omega)$
leads to massive improvement of signal-to-noise ratio and boosts up
the separation estimation precision. 

\subsection*{Protocol}
\begin{figure*}
\includegraphics[width=1\textwidth]{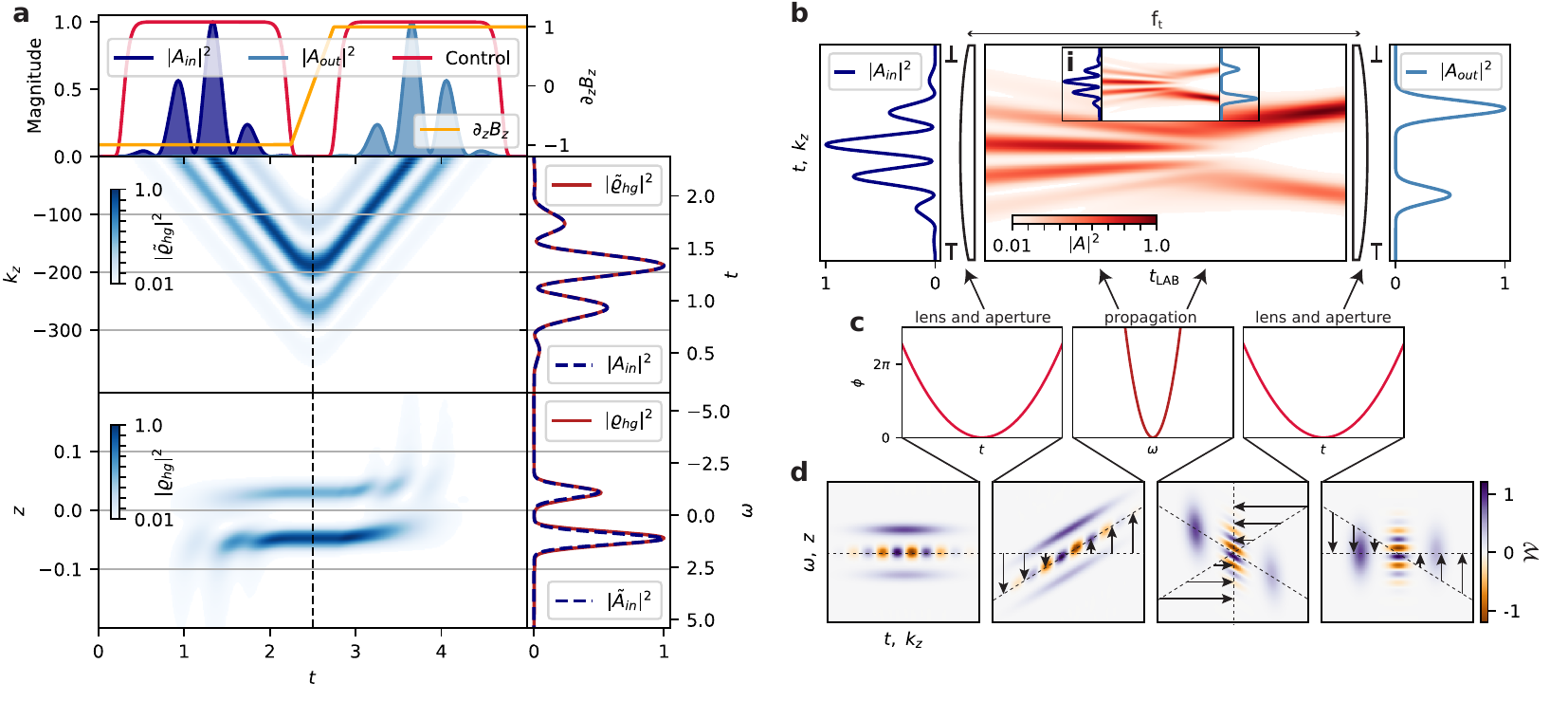}\caption{\textbf{Temporal imaging in GEM.} \textbf{a} Simulation of GEM protocol for signal ($A_{\mathrm{in}}$) with two spectral components with finite bandwidth. The maps show evolution of the atomic coherence ($\rho_{\mathrm{hg}}$) in the wavevector (top) and real space (bottom). The frequency-to-space and time-to-wavevector mapping feature of GEM protocol is visible on the crossections (taken along the dashed lines) of the coherence maps that are compared to the temporal and spectral shape of the stored optical field. \textbf{b} Temporal far field imaging using two time-lenses separated by temporal propagation. The output signal is a Fourier transform of the input signal. The temporal aperture clips the input signal and sets the frequency resolution limit. The inset \textbf{i} shows propagation of the same input signal through similar TI setup but with a lens with negative focal length followed by backward temporal propagation, resulting in the inverted spectrum at the output. \textbf{c} Phase shapes that are sequentially imprinted on the input signal to implement the lens-propagation-lens TI setup. \textbf{d} Time-frequency and space-wavevector Wigner representation ($\mathcal{W}$) of the input signal at each stage of the TI setup. The arrows represents the transformations of the phase-space upon each transformation (lens or propagation).\label{fig:TI_GEM}}
\end{figure*}

The idea for achieving sub-Fourier performance in our spectrometer
via the PuDTAI protocol relies on engineering a measurement that
can split the signal pulse into symmetric and antisymmetric combinations
with respect to the mean time or frequency. The protocol that realizes
this projective measurement is inspired by a technique known from
conventional (real space) super-resolved imaging called SLIVER (superlocalization
via image-inversion interferometry) \citep{Wicker2007,Wicker2009,Nair2016,Larson2019,Nair2016a},
where an image is inverted in one arm of a Mach-Zhender interferometer. The Fisher information
associated with photon detection probabilities in the symmetric $\mathcal{P}_{\hgz}$
and antisymmetric $\mathcal{P}_{\hgo}$ ports exhibits extraordinary
sensitivity for small ($\varepsilon\ll1$) separation estimation $\mathcal{F}_{\mathrm{SLIVER}}\approx\frac{1}{4}-\frac{\varepsilon^{2}}{32}$
and for $\varepsilon\to0$ approaches $\mathcal{F}_{\mathrm{Q}}$.

In our protocol, schematically depicted in Fig.~\ref{fig:pudtai}
we divide the signal in half, rather than splitting the signal to
prepare two identical copies. This method while being slightly different
from SLIVER, achieves the same sensitivity to source separation and
inherently, as an additional feature, allows use with $N$-photon
states \citep{Larson2020}.
This framework assumes prior knowledge of the mean frequency, also
known as the source centroid. This assumption is practically valid,
as an adaptive strategy can be used where the centroid is first estimated
with desired precision using direct detection \citep{Grace:20}.

\subsection*{Temporal Imaging in GEM}

\begin{figure}
\includegraphics[width=\columnwidth]{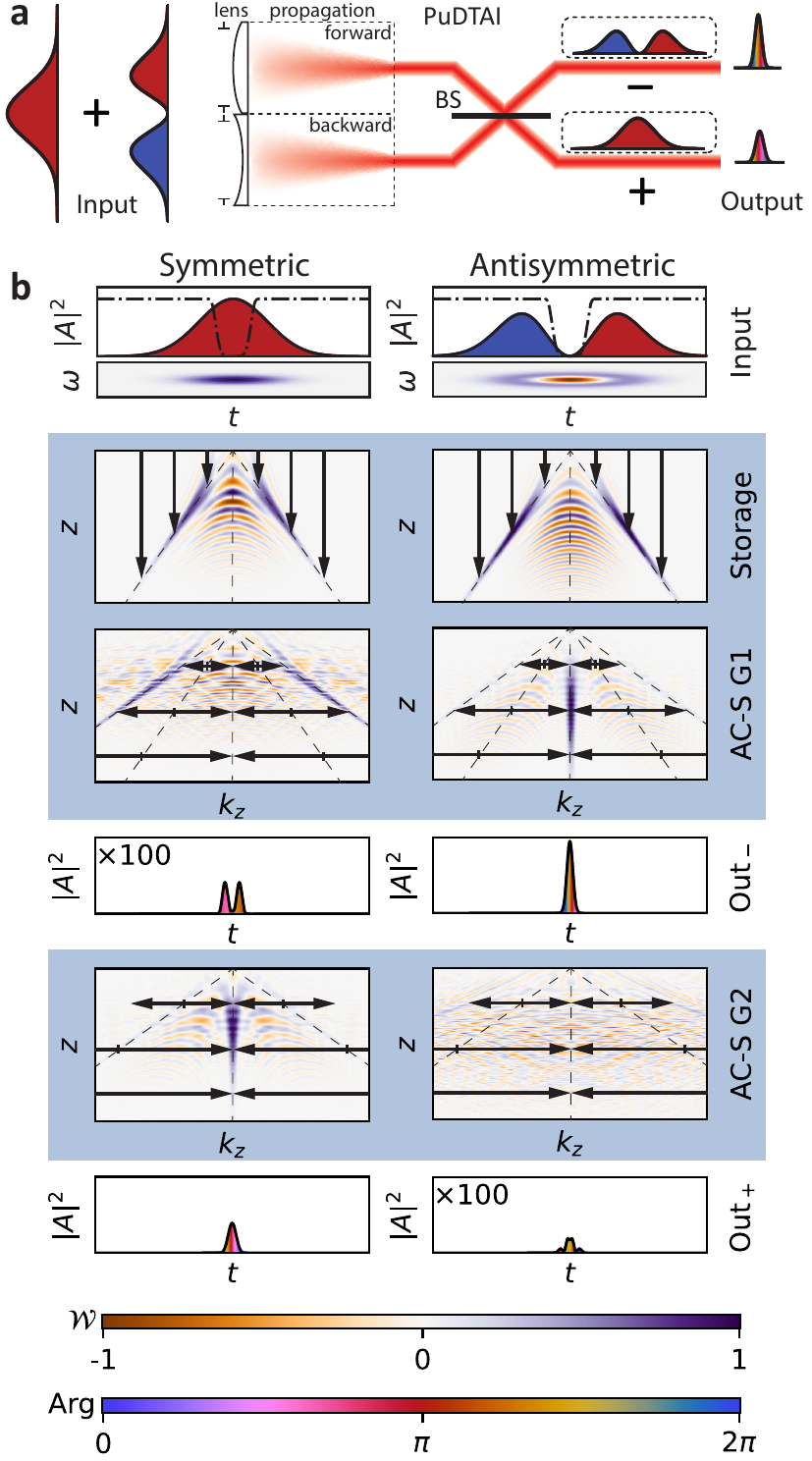}\caption{\textbf{Step-by-step realization of the protocol for symmetric and
antisymmetric input signals. a} Schematic representation of the \sasin\
interferometer superimposing the divided signal in the Fourier space. The color fill represents the phase (Arg) of the signal field envelope, BS - beamsplitter.
\textbf{b } The corresponding spin-wave evolution in $(z,k_{z})$ space
is represented by Wigner function maps ($\mathcal{W}$). The time-frequency domain Wigner maps are shown for reference. Thanks to the symmetrically-chirped
control field used in GEM protocol (Storage) the two halves of the
signal pulse are mapped in a symmetric manner on separate Fourier
($k_{z}$) components while overlapping in the longitudinal coordinate
$z$. The black dashed lines represent the temporal aperture that removes the central part of the singnal pulse during the mapping. The spatially-resolved AC-Stark phase modulation (AC-S) with
chirped square grating splits the spin-waves in the $k_{z}$ direction
and makes the two halves to interfere at central $k_{z}$ coordinate
- the antisymmetric port of the interferometer which is then read-out
by control field pulse (Out$_{-}$). The second (symmetric) port is
restored by modulating the spin waves again with a second chirped
square wave grating with opposite phase (AC-S). The symmetric port
is finally retrieved (Out$_{+}$). \label{fig:Step-by-step}}
\end{figure}

Before introducing the details of the \sasin\ protocol, let us first discuss a simpler case of DI spectrometer based on an atomic Gradient Echo Memory (GEM) \citep{Mazelanik2020,Cho2016,Hosseini2009} that is realized in the same GEM setup schematically depicted in Fig. \ref{fig:Experimental-setup}\textbf{a}. In this setup a   
signal pulse with a slowly varying amplitude $\mathcal{A}(t)$ enters the atomic cloud placed in the magnetic field gradient along the propagation axis $g=\partial_zB_z$. At the same time the cloud is illuminated by a strong, co-propagating control field $\mathcal{C}$ that due to two-photon interaction maps the signal field onto atomic coherence described by the off diagonal terms $\varrho_{hg}(z)$ of the spatially dependent $\{|g\rangle,|h\rangle\}$ two-level atom density matrix $\hat{\varrho}(z)$. To restore the signal light, the gradient is switched to the opposite $g\to-g$ and the control field pulse is applied again, as shown in the corresponding sequence in Fig. \ref{fig:Experimental-setup}\textbf{b}. In Fig. \ref{fig:TI_GEM}\textbf{a} we show numerical simulation of the GEM protocol for a signal pulses with two spectral components of different amplitudes. The density maps below the time trace show the evolution of the atomic coherence in the real space and wave vector coordinates. The presence of the magnetic field gradient enables the 
crucial feature of the GEM protocol which is the spectro-spatial mapping that links spectral
components of the signal pulse with the amplitude of the atomic coherence
along the ensemble $\mathcal{\tilde{A}}(\omega)\leftrightarrow\varrho_{hg}(z)$.
At the same time, in time domain, the pulse shape is transferred to
wavevector-space components of the coherence $\mathcal{A}(t)\leftrightarrow\tilde{\rho}_{hg}(k_{z})$, where $\tilde{\varrho}_{hg}(k_{z})$ is the Fourier transform of the real space coherence. This correspondence is directly visible on the cross sections of the atomic coherence maps (right panels) taken along the dashed lines where we also show the optical field magnitudes in the time and spectral domain.   

The DI imaging spectrometer utilizes a temporal imaging (TI) technique to perform a Fourier transform of the signal pulse and thus map a spectrum of the signal pulse onto temporal shape, as illustrated in Fig.~\ref{fig:TI_GEM}\textbf{b}. This is achieved by applying sequentially temporal and spectral phase modulations corresponding to lens-propagation-lens components. A time lens corresponds to applying a quadratic phase in time, while propagation is achieved by imprinting a quadratic phase in spectrum onto the signal pulse as shown in Fig. \ref{fig:TI_GEM}\textbf{c} \citep{Karpinski2017}. In GEM, the temporal phase modulation is achieved by chirping the control field that controls the light-coherence mapping process. The temporal phase modulation is this way reflected in the phase of the coherence in the wavevector space ($k_z$). On the other hand the phase modulation in the spectral domain is achieved by Spatial Spin-wave Modulation (SSM) \citep{Parniak2019,Mazelanik2019,Lipka2019}, that utilizes spatially varying ac-Stark shifts to imprint arbitrary phase profiles onto the coherence in the real space coordinate ($z$). The ac-Stark shifts are caused by an intensity-shaped beam that illuminates the atoms from the side (see Fig. \ref{fig:Experimental-setup} for the geometry and the actual sequence of the experiment) during the storage time.  

The most convenient way to understand evolution of the optical field or the corresponding atomic coherence
at each step of the protocol is to investigate a phase-space quasi-probability
given by the Wigner function 
\begin{equation}
\mathcal{W}_\mathcal{Q}(q, p)=\frac{1}{\sqrt{2\pi}}\int \mathcal{Q}(q+\xi/2)\mathcal{Q}^{*}(q-\xi/2)\exp(-ip\xi)\mathrm{d}\xi\label{eq:wigner}
\end{equation}
 where the quantity of interest $\mathcal{Q}$ and the conjugate variables $\{q, p\}$ are either $\mathcal{A},\ \{t, \omega\}$ for which it becomes a chronocyclic Wigner function
\citep{Cohen1995} describing the signal pulse or $\varrho_{hg},\ \{k_z, z\}$ for the atomic coherence.

In Fig. \ref{fig:TI_GEM}\textbf{d} we show evolution of the chronocyclic Wigner function for the input signal propagating through the TI setup from Fig. \ref{fig:TI_GEM}\textbf{b}. A temporal phase modulations $\mathcal{A}(t)\to\mathcal{A}(t)\exp\left(i\phi(t)\right)$  
reshapes the chronocyclic Wigner function $\mathcal{W}_\mathcal{A}(t, \omega)$ in the $\omega$ direction, at the same time the corresponding coherence quasiprobability is reshaped in the $z$-direction $\mathcal{W}(k_{z}, z)\xrightarrow{\phi(t)}\mathcal{W}(k_{z}, z')$. For example, in Fig. \ref{fig:TI_GEM}\textbf{d} the quadratic phase profile that implements the time-lens stretches the $\omega$ axis on the $\mathcal{W}_\mathcal{A}(t, \omega)$ map linearly with $t$ as marked by the arrows. 

The spectro-spatial mapping enables implementation of spectral phase
operations by phase-modulating the coherence in real-space coordinates
$\varrho_{hg}(z)\to\varrho_{hg}(z)\exp(i\chi(z))$. These are linked to $k_{z}$-direction reshaping of the quasiprobability $\mathcal{W}(k_{z}, z)\xrightarrow{\chi(z)}\mathcal{W}(k_{z}', z$) and $t$-axis transformation of the chronocyclic Wigner function. This is visible in the third panel of Fig. \ref{fig:TI_GEM}\textbf{d} corresponding to propagation transformation realized by a quadratic spectral phase profile. In the last panel, the second time-lens operation completes the $90^\circ$ rotation of the phase - the Fourier transform of the input signal amplitude $\mathcal{A}(t)$ is complete. In practice, however one may omit the last step as it only corrects the phase of the resulting output signal that is (the phase) irrelevant in case of phase insensitive detection as photon counting.

The discussed TI setup not only serves as introductory example but is implemented experimentally in our memory \citep{Mazelanik2020} to serve as DI reference to our superresolution protocol which we describe below. 

\subsection*{\sasin\ interferometer}
Finally, within the introduced framework, we are able to realize both the time-dependent-transmission
mirror that divides the pulse and performs a time-inversion as well
as beamsplitter that combines the two components and performs the
final projection. The sequence of operations performed using the quantum memory is shown in Fig. \ref{fig:Experimental-setup}\textbf{b}.

The step-by-step evolution of the quasiprobability
$\mathcal{W}(z,k_{z})$ in the \sasin\ protocol for two input pulse shapes
$\psi_{\hgz}$ and $\psi_{\hgo}$ is presented in Fig. \ref{fig:Step-by-step}\textbf{b}.
First, similarly to TI,  a temporal phase modulation with $\phi(t)=-\alpha t|t|/2$, corresponding
to $z\to z'=z+\alpha|k_{z}|$ transformation in the phase space implements
the time-dependent-transmission beamsplitter and simultaneously performs
the time-inversion of the first part. In fact, such modulation represents a dual time-lens that is half convex, half concave i.e. the  focal length of the first half ($t<0$) is positive when for the second part ($t>0$) is negative. The Wigner function maps (dark
background) illustrate that the two parts of the pulse are mapped symmetrically on the opposite sides of the $k_{z}$
space. For practical reasons the temporal phase modulation is accompanied with Cassegrain-type temporal aperture to prevent mapping of the central part of the pulse (around $t=0$), the aperture is represented by a black dashed line on the signal pulses time traces (first row of Fig. \ref{fig:Step-by-step}\textbf{b}). The (minor) influence of the aperture on the protocol's performance is discussed in the Methods section.

Next, by performing $k_{z}$-direction splitting transformation in the phase space we
overlap the two parts and make them interfere. This is achieved by
imprinting a $\pi$-depth square-wave grating (G1) with linearly increasing
grating wavevector $k_{g}=\kappa z$ (see Fig.~\ref{fig:Experimental-setup}\textbf{b}).
The chirped grating transfers the spin wave into diffraction
orders with $z$-dependent wavevector spacing $k_{z}\to k_{z}'=k_{z}\pm\kappa z$.  Let us now focus on the relevant diffraction orders: the minus-first in the negative time part and the first one in the positive time part. Those two orders in the language of temporal imaging are corresponding to forward and backward propagation respectively. 
If we combine this with the dual-lens operation implemented in the first step we achieve dual-far-field temporal imaging setup that performs forward Fourier transform of the first half of the pulse and backward Fourier transform of the second half. Moreover, as the resulting parts overlap in the Fourier domain they interfere as represented in Fig. \ref{fig:Step-by-step} \textbf{a}.      
For this to happen the propagation distance must be equal the focal length of the time lens, the condition is met for the grating
chirp parameter $\kappa$ equal to the inverse of the storage chirp $\kappa=\nicefrac{1}{\alpha}$.
The third row of Fig.~\ref{fig:Step-by-step}\textbf{b} represents the Wigner
function after application of the grating.  Here we see that the two signal pulse components are forced by the grating (G1) to interfere at $k_{z}=0$ coordinate
(the symmetry axis of the quasiprobalitlity distribution).  By taking a $z$-direction
integral of the quasiprobability around $k_{z}=0$ we retrieve light intensity at the
output of the memory. Moreover, the phase between the two interfering components can be adjusted by shifting the grating phase by $\zeta$: $\mathrm{sq}(\kappa z^{2})\to\mathrm{sq}(\kappa z^{2}+\zeta)$.  
For $\zeta$=0 the interference for symmetric input shapes as $\mathcal{G}$  is destructive, while for antisymmetric
$\mathcal{HG}$ it becomes constructive, as can be seen in Fig. \ref{fig:Step-by-step}\textbf{a}. To access the symmetric port the coherence is modulated again with
similar square-wave grating (G2), but with doubled chirp parameter $k_{g}=2\kappa z=\frac{2}{\alpha}z$,
and shifted in phase by $\pi/2$. With this modulation we simply reverse previous temporal propagation step, and make it again but with opposite phase between the two components. This is illustrated in the fifth
row of Fig. \ref{fig:Step-by-step}, where we observe constructive
interference for the symmetric input shape, and destructive for antisymmetric
one. Finally, the area around $k_{z}=0$ is remapped to light which
constitutes the output from the symmetric port of the interferometer
(last row of Fig. \ref{fig:Step-by-step}). Notably, the Wigner space domain transformations are the same for both input shapes, and the different results at the output are caused solely by the difference in the input shape. In Table \ref{tab:summary} we provide a survey of the phase operations needed to implement the \sasin\ protocol. 
To summarize, the protocol realizes a $\pi/2$ ($-\pi/2$) rotation of the Wigner function of the first (second) half oh the signal pulse, and interferes the two parts in the Fourier domain. Additionally, as at the output have the the Fourier domain mapped to the time axis, a longer signal pulses give shorter pulses
at the output of the interferometer that leads to higher signal-to-noise
ratio when using noisy detectors. This provides an additional practical
advantage while dealing with very narrowband states of light.  
\begin{table}
\centering
\arrayrulecolor[rgb]{0.8,0.8,0.8}
\begin{tabular}{|c|c|c|c|} 
	\hline
	Function                                                           & Type                     & Formula                             & Wigner space                  \\ 
	\arrayrulecolor{black}\hline
	lens                                                               & $t\leftrightarrow k_{z}$ & $\phi(t)=\frac{\alpha}{2} t^{2}$    & $z\to z-\alpha k_{z}$         \\ 
	\arrayrulecolor[rgb]{0.8,0.8,0.8}\hline
	propagation                                                        & $z$                      & $\chi(z)=\frac{\beta}{2} z^{2}$     & $k_{z}\to k_{z}+\beta z$      \\ 
	\hline
	dual lens                                                          & $t\leftrightarrow k_{z}$ & $\phi(t)=-\frac{\alpha}{2} t|t|$     & $z\to z +\alpha |k_{z}|$        \\ 
	\hline
	\begin{tabular}[c]{@{}c@{}}bidirectional\\propagation\end{tabular} & $z$                      & $\chi(z)=\mathrm{sq}(\kappa z^{2})$ & $k_{z}\to k_{z}\pm \kappa z$  \\
	\hline
\end{tabular}
\arrayrulecolor{black}
\caption{Summary of the temporal and spectral phase modulations used to implement both \sasin\ and DI spectrometer. The $t \leftrightarrow k_{z}$ type of transformation means that it is implemented during the light-to-atoms mapping stage by temporal phase modulation of the control field, while the $z$ stands for the real-space phase modulation of the atomic coherence. The formula for the square wave pattern is $\mathrm{sq}(\xi)=\pi \left((-1)^{\left \lfloor{\xi/\pi}\right \rfloor}+1\right)/2$. Each modulation has its corresponding Wigner space transformation that can be expressed in the $\{k_{z}, z\}$ or $\{t, \omega\}$ coordinates.\label{tab:summary}}
\end{table}

\subsection*{Experimental Results}

\begin{figure*}
\includegraphics[width=1\textwidth]{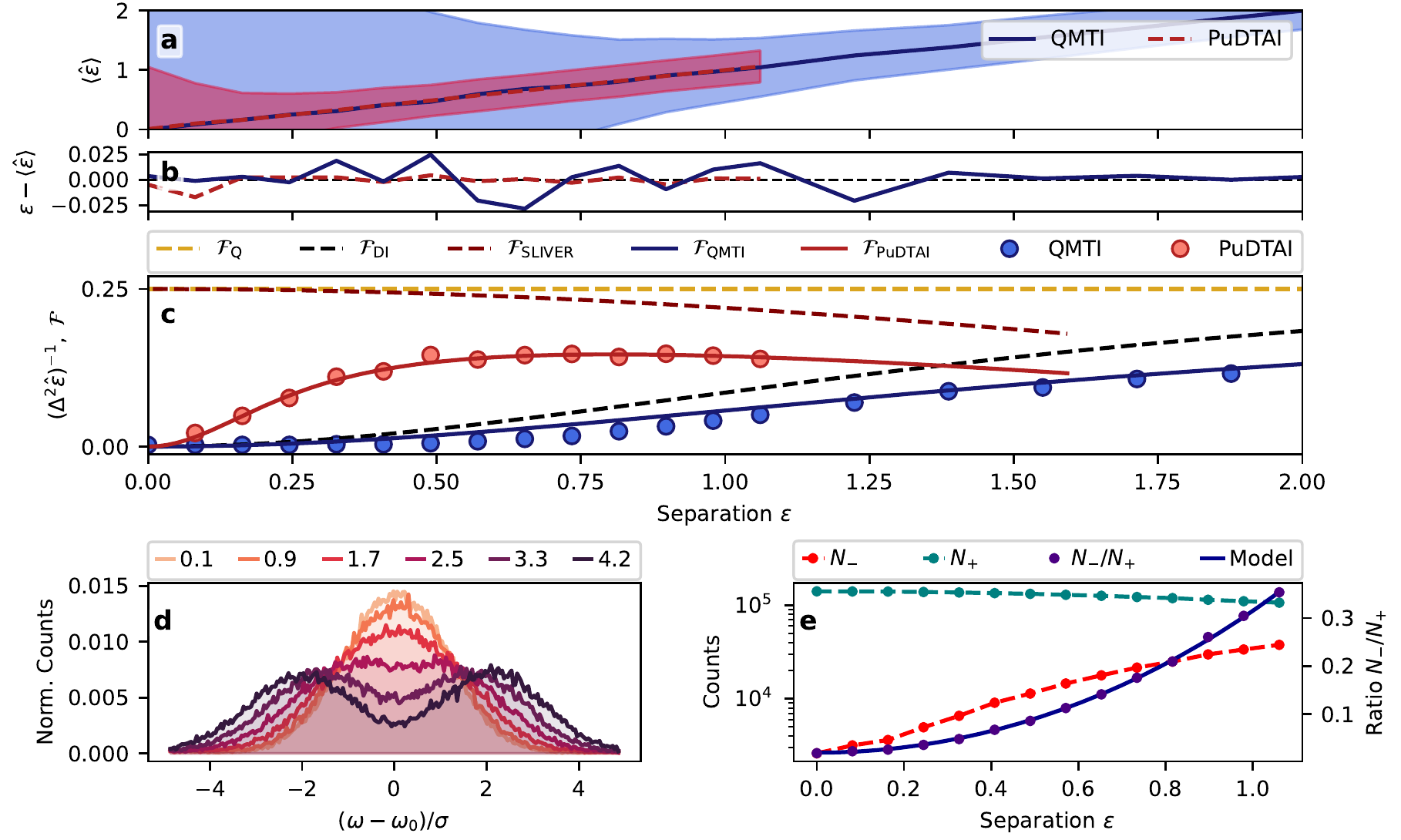}\caption{\textbf{Separation estimation using \swft\  and \sasin\  protocols.}
\textbf{a} Raw $\varepsilon$ estimates obtained from DI approach
(\swft\ ) and \sasin\  protocol. The shaded regions represent estimation
uncertainty given by the square root of the estimator variance (SEM)
normalized for 10 processed photons. \textbf{b} Estimator's biases
for both schemes. \textbf{c} Estimation precision compared with idealized
and real world CRBs given by corresponding Fisher information ($\mathcal{F}_{\mathrm{Q}}$-
Quantum Fisher information, $\mathcal{F}_{\mathrm{SLIVER}}$ - ideal
SLIVER protocol, $\mathcal{F}_{\mathrm{DI}}$ - ideal DI spectrometer,
$\mathcal{F}_{\mathrm{\swft}}$ - Quantum memory temporal imaging,
$\mathcal{F}_{\mathrm{\sasin}}$ - \sasin\  protocol). \textbf{d}
Frequency-labeled single photon counts distributions obtained with
\swft\ spectrometer for different true separation values $\varepsilon$.
\textbf{e} Contributions to the antisymmetric ($N_{-}$) and symmetric
($N_{+}$) port of the \sasin\  device and the ratio $N_{-}/N_{+}$
along with calibrated theoretical model. \label{fig:-Separation-estimation}}
\end{figure*}

To benchmark our protocol we artificially prepare signal pulses composed
of two mutually incoherent spectral components 
\begin{equation}
\mathcal{\tilde{S}}_{\varphi}(\omega)=\frac{1}{\sqrt{2}}\left(\tilde{\psi}_{\hgz}(\omega-\sigma\varepsilon/2)+e^{i\varphi}\tilde{\psi}_{\hgz}(\omega+\sigma\varepsilon/2)\right)\label{eq:signal}
\end{equation}
where the random phase $\varphi\in[0,2\pi[$ is drawn from uniform
distribution. The corresponding chronocyclic Wigner function \citep{Cohen1995}
of such signal distribution reads: 
\[
\mathcal{W}_{\mathcal{S}}(t,\omega)=\psi_{\hgz}^{2}(t)(\tilde{\psi}_{\hgz}^{2}(\omega-\sigma\varepsilon/2)+\tilde{\psi}_{\hgz}^{2}(\omega+\sigma\varepsilon/2))
\]
We send the pulses to the memory where they are processed by our pulse-division
time-inversion interferometer. To read-out the contributions in symmetric
and antisymmetric ports we use two sequentially applied pulses of
the control field that are interleaved with the second AC-Stark modulation.
The processed signal light coming from the memory is detected using
single-photon counting module (SPCM) and photo-counts are time-tagged
to identify the output ports (see Fig. \ref{fig:porty}\textbf{a} in the Methods section for the time-bin
histogram). The mean single-shot signal contribution summed over the
symmetric and antisymmetric port was set to be around $\bar{n}\approx0.69$.
After many experimental repetitions we count the total contributions
$N_{-}$, $N_{+}$ in the antisymmetric and symmetric port respectively
and calibrate the maximum-likelihood estimator $\hat{\varepsilon}(N_{-}/N_{+})$
based on theoretically expected counts ratio $N_{-}/N_{+}$. The estimator
is then used to estimate the value of the separation $\varepsilon$. For this, for each $\varepsilon$ value we collected approx. $1.5\times10^{5}$ counts. To estimate the estimator variance we follow the bootstrapping technique: 
for each $\varepsilon$ setting we randomly prepare $10^{3}$ sets of samples, each containing $1.5\times10^{5}$ total
counts. The estimator is then evaluated on each set and the mean and
variance of $\hat{\varepsilon}$ is calculated. In Fig. \ref{fig:-Separation-estimation}\textbf{a}
we show raw estimation results for \sasin\  protocol and DI approach
on a common $\langle\hat{\varepsilon}\rangle$ plot. The filled regions
corresponds to the estimation uncertainty given by the square root
of the estimator variance and normalized to 10 processed photons. Both methods give similar values of the separation parameter, but as we expect the \sasin\ protocol vastly outperforms the DI over the whole $\varepsilon<1$ range. The improvement expressed in variance ratio for the same experimental conditions is most prominent for small separation values and reaches about 20 for $\varepsilon=0.08$. In other words using \sasin\ protocol we need, 20 times less photons to achieve the same precision. 

The measurements in DI scheme are obtained by using ultranarrowband
far-field temporal imaging technique which we call quantum-memory
temporal imaging (\swft) \citep{Mazelanik2020}. The protocol is implemented in
the same GEM device following the recipe from the introductory section.  It performs a $90^{\circ}$ rotation of the signal pulse Wigner space which results in spectrum-to-time
mapping $\omega\to2\alpha_{\mathrm{DI}}t$, where $\alpha_{\mathrm{DI}}$
is the control field chirp parameter corresponding to the time-lens
magnitude. The output signal is detected using the same SPCM as in
the case of \sasin, where counts corresponding to distinct spectral
components have different arrival time (timetags). The counts distributions
with already calibrated frequency labels are presented in Fig. \ref{fig:-Separation-estimation}\textbf{d}.
From those, we estimate the separation using maximum-likelihood separation
estimator $\hat{\varepsilon}_{\mathrm{DI}}$. The variance of this
estimator is calculated by following the same bootstrapping technique
as in the case of \sasin\  protocol. Figure \ref{fig:-Separation-estimation}\textbf{b}
portrays the estimators' biases $\langle\hat{\varepsilon}\rangle-\varepsilon$
proving unbiasedness of both estimators even for small $\varepsilon$
and validating the results for variances.

In Fig. \ref{fig:-Separation-estimation}c we compare the achieved
precision with CRB for both schemes. Here we also show the ultimate
bound set by the QFI ($\mathcal{F}_{\mathrm{Q}}$) as well as bounds
for ideal SLIVER protocol ($\mathcal{F}_{\mathrm{SLIVER}}$) and idealized
DI ($\mathcal{F}_{\mathrm{DI}}$). The real bounds are placed by FIs
that take into account experimental imperfection such as detection
noise and finite bandwidth of the ensemble - $\mathcal{F}_{\mathrm{\swft}}$,
as well as finite interferometer visibility and interferometer losses
in the case of \sasin\  protocol - $\mathcal{F}_{\mathrm{\sasin}}$
(see \methods). We see that the \sasin\  protocol vastly outperforms
the QMTI for normalized separations $\varepsilon<1$ with maximum
improvement in terms of variance of about 30 for $\varepsilon=0.4$. Finally,
we show the data used for estimation, to better illustrate the origin
of sensitivity enhancement. In Fig.~\ref{fig:-Separation-estimation}\textbf{d}
the results of direct imaging are presented, and we observe that for
$\varepsilon<1$ the shape of the obtained spectrum remains close
to a Gaussian. On the other hand, in Fig. \ref{fig:-Separation-estimation}\textbf{e}
we show total contributions to the antisymmetric ($N_{-}$) and symmetric
($N_{+}$) ports, as well as the ratio $N_{-}/N_{+}$ along with fitted
model that is used to estimate the $\varepsilon$ without a direct
need to know total source brightness.

\section*{Discussion}

\begin{figure}
\includegraphics[width=1\columnwidth]{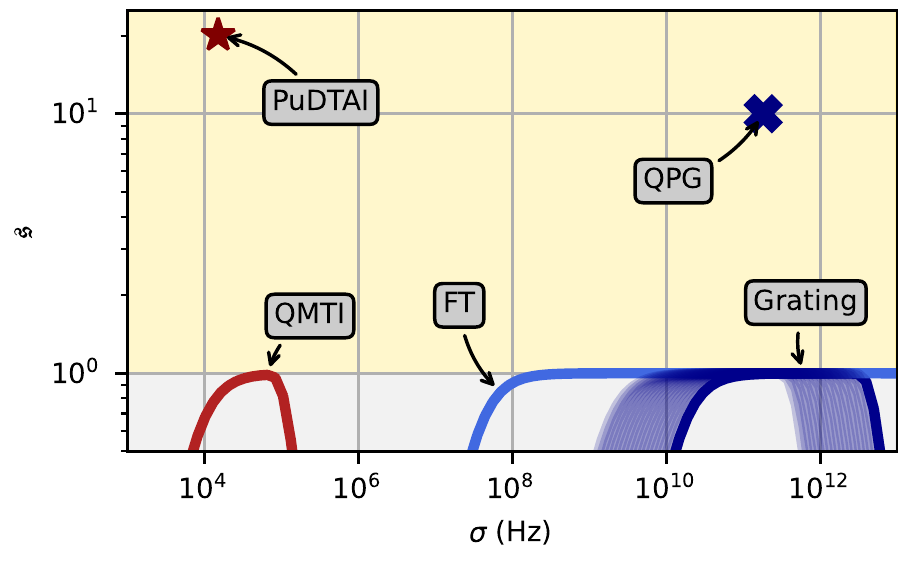}\caption{\textbf{Comparison of super-resolution enhancement factor $\mathfrak{s}$
of different spectrometers and super-resolution spectroscopy techniques.}
All conventional methods (Grating, FTS - Fourier transform spectrometer)
and the temporal-imaging method using our quantum memory (QMTI - quantum-memory
temporal imaging) are analogous to the DI approach and thus fall under
the Fourier limit. Only two (QPG - Quantum Pulse Gate \citep{Donohue2018}
and \sasin\  - this work) beat the limit of $\mathfrak{s}=1$
and thus achieve super-resolution. The curves represent exemplary
realizations of given spectrometer for constant time-bandwidth
product. \label{fig:Comparison}}
\end{figure}

We have demonstrated frequency-separation estimation which outperforms
direct spectroscopy. The protocol operates in a previously unexplored
regime of very narrowband light, which merits further discussion and
comparison with other possible schemes. In Fig.~\ref{fig:Comparison}
we compare several different approaches to measure frequency difference
of two sources. We characterize the gain achieved with our method
via the super-resolution parameter $\mathfrak{s}$ that is interpreted
as a reduction of resources (number of photons) required to achieve the same
resolution as direct imaging spectroscopy with the same aperture
$\sigma$. The $\mathfrak{s}$ for a given measurement scheme characterized
by the Fisher information $\mathcal{F}$ is calculated as 
\begin{equation}
\label{eq:limitS}
\mathfrak{s}=\underset{\varepsilon\to0}{\lim}(\mathcal{F}/\mathcal{F}_{\mathrm{DI}}).
\end{equation}
While for the perfect case with $\mathcal{F}=\mathcal{F}_Q$ the superresolution enhancement would reach infinity, this is practically never the case. In particular, it has been recently shown that effects such as finite visibility, noise, or cross-talk will always restore the $\varepsilon^2$ scaling \cite{PhysRevLett.126.120502}, making the above limit well-defined. For instance, in the case of SLIVER with finite visibility we will have:
\begin{equation}
\mathfrak{s}_\mathrm{SLIVER}= \frac{\mathcal{V}^2}{2(1-\mathcal{V}^2)}.
\label{eq:sliverV}
\end{equation}
Thus, for a high visibility the enhancement may be immense, yet remains well-defined for small $\varepsilon$. The \sasin\ protocol follows a similar scaling, with the full formula given by Eq. \ref{eq:sasinS} in the Methods section that yields $\mathfrak{s}=20\pm 0.5$.

All the conventional methods that include: grating and Fourier transform
(FT) spectrometers along with far-field temporal imaging or inherently
lossy scanning methods that employ cavity or EIT media fall into DI
description and are Fourier-limited ($\mathfrak{s}=1$).
Surprisingly, there is only one more demonstration (Quantum Pulse
Gate – QPG \citep{Donohue2018}) which yielded improvement over the conventional
DI approach. Until now, the QPG approach, that in principle enables temporal-mode
demultiplexing, allowed projecting only a single mode at a time. In
our work we are able to observe two modes (ports) in the same experiment.
An important future challenge in both techniques is to allow truly multi-mode
sorting.

While using DI (a spectrometer) for high bandwidth signals seems an
obvious choice, one may conclude that for narrowband signals a balanced
heterodyne (or homodyne) detection would perform better. The first
obvious challenge is that one then needs a stable and narrowband local
oscillator, but as our method uses a control field, this requirement
is effectively shared. However, when the sources brightness is low
(below one photon per mode), the heterodyne detection scheme suffers
strongly from shot noise. For the particular task of separation estimation,
the shot noise will affect the estimation precision and makes it useless
for measuring single-photon-level sources. Nevertheless, heterodyne
(or homodyne) can overcome the Fourier limit for higher signal photon
levels and for small separations as has been recently analyzed in
detail for the spatial domain \citep{Datta2020} ($\mathfrak{s}>1$
for average 7.2 photons per mode and more), with the use of optimized
data analysis, as opposed to simply looking at power spectral densities.
Therefore, a purely photon-counting method is in general needed in
the low light-level regime.

It is also worth considering the influence of fluctuations on our method. If the amplitude or phase of the signal varies over the single shot (temporal aperture), clearly the two parts of the pulse will not be able to interfere perfectly. Such a situation represents and unavoidable loss of information, and may be quantified via the reduced visibility $\mathcal{V}$ of interference. The result is therefore the same as already considered in the case of finite visibility caused by the device itself (see Eq. \ref{eq:sliverV} and Methods), yet the information here is already reduced in the signal itself.

The conjunction of the operation in the optical domain and the kHz-level
resolution has promising application in the schemes that combine microwave
or radio-frequency and optical domains \citep{Mirhosseini2020}. Our
method of spectroscopy can provide an optimal detection scheme for
light that is transduced from the RF or microwave domain, without
the sensitivity loss due to shot noise, which is inherently present
in the traditional heterodyne optical receivers.

We envisage that other platforms may be used to implement the protocol
we proposed, also in different frequency regimes. For example, rare-earth
doped crystals such as Eu:Y$_{2}$SiO$_{5}$ offer narrow homogeneous
absorption linewidth that can be dynamically broadened with an electric
field \citep{PhysRevLett.96.043602} for the purpose of time-frequency
domain multimode storage \citep{doi:10.1021/acs.nanolett.0c02200}.
Processing with AC-Stark shifts has also been demonstrated \citep{Yang2018},
and advances in embedding ions in waveguides promise good efficiency
and noise properties.

Another approach has been very recently discussed theoretically by
Shah and Fan \citep{Shah2021}, who considered a transformation
device based on pulse shapers and electro-optic modulators. This approach
may be very applicable where relevant devices can operate, which is
in particular GHz-bandwidth regime in the telecom band of light. In
this regime no quantum memory is required, as the timescales involved
are much shorter. Several other quantum-information protocols have
been realized with similar means, also for quantum light, and the
scheme is generally termed quantum-frequency processor \citep{Lukens2017}. 
We expect that those approaches would perform particularly well in GHz-bandwidth regime and beyond. On the other hand, to enhance resolution of our scheme to below 1 kHz a quantum memory with much longer lifetime would suffice, which has already been demonstrated in an optical lattice \cite{PhysRevA.87.031801}. This lifetime does not necessarily change the superresolution enhancement factor itself, which for the given temporal/spectral aperture only depends on the quality of interference and noise properties of the device.

Remarkably, the \sasin\  protocol has been realized here in a quantum
memory that is insensitive to the signal spatial distribution., i.e. is spatially-multimode \cite{Parniak2017}. This means
that it can accept external fluorescence light that is in principle
spatially-multimode. Furthermore, we note that to analyze fluorescent
light from different kinds of samples, one  needs to match the wavelength
of operation. This can be done via the maturing techniques of quantum
frequency conversion \citep{Maring:18}.

The path from resolving two sources to obtaining a more complex spectrum
with super-resolved features is challenging, yet some knowledge can
be drawn from imaging problems considered so far. Various groups have
considered the cases of three sources in two dimensions \citep{PhysRevA.99.013808,PhysRevA.99.012305},
or two sources with arbitrary brightness \citep{PhysRevA.98.012103},
and practical frameworks for multi-pixel images are being developed
as well \citep{2105.01743}. For the spectroscopic case discussed
here, different operations in the quantum memory will be needed to
prepare measurements optimal for multi-source scenarios.

\section*{Methods}

\subsubsection*{Gradient Echo Memory}

The Gradient Echo Memory is based on pencil-shaped ($8\times0.3\times0.3\,\mathrm{mm}^{3}$)
$\mathrm{^{87}Rb}$ atomic cloud prepared in magnetooptical trap.
The full experimental sequence is depicted in Fig. \ref{fig:Experimental-timing-sequence}\textbf{a}.
After the cloud preparation stage (trapping, compression and cooling)
the atoms are optically pumped to the $|h\rangle=(F=2,m_{F}=-2)$
state. The control field (795 nm) couples the excited state $|e\rangle=(F=1,m_{F}=-1)$
with the previously emptied storage state $|g\rangle=(F=1,m_{F}=0)$
which enables coherent absorption of the signal field at the $|h\rangle\to|e\rangle$
transition (see \ref{fig:Experimental-timing-sequence}\textbf{b} for reference). The magnetic field gradient crucial to the GEM protocol
is generated by two identical coils located antisymmetrically on
the $z$-axis and powered by fast H-topology current switch, that
allows switching between negative and positive gradients within 5
$\mu$s. In the experiment we set the gradient to $\partial_z B_z = \pm7.3\ \mathrm{\mu}\mathrm{{T}}/\mathrm{mm}$
Additionally, to separate the magnetically broadened absorption spectrum
of the $|h\rangle\to|e\rangle\to|g\rangle$ transition from the unbroadened
clock transition $(F=2,m_{F}=-1)\to(F=1,m_{F}=0)\to(F=1,m_{F}=1)$
we keep the cloud in bias magnetic field of $B_z=120$~$\mu$T magnitude
along the $z$-axis. Moreover, to stabilize the two photon detuning
that is sensitive to magnetic fields we operate the protocol at 50
Hz synced with local mains frequency and correct for slow magnetic
fluctuations (caused, for example, by an elevator near our laboratory)
by adjusting the magnitude of the bias magnetic field in a feedback
scheme described below.
The temperature of the cloud is measured to be about $60\,\mu \mathrm{K}$. Given the configuration of laser beams, i.e. angle between coupling and signal light, the atomic coherence has a $50\,\mathrm{rad/mm^{-1}}$ transverse wavevector component. This gives $\approx 260\,\mu \mathrm{s}$ of thermal-motion-limited storage time \citep{Lipka2021}. The control field induced spin-wave decay rate \citep{Mazelanik2020} in the experiment is about $10\,\mathrm{kHz}$.

The signal photons are passed through an optically-pump $^{87}$Rb-vapour
cell in order to remove the residual control field. At the end, they
are coupled to a single-mode fiber and detected with an SPCM.

\subsubsection*{Magnetic field synchronization and stabilization}

The protocol operates at 50 Hz repetition rate synced with local mains
frequency. The synchronization mechanism is realized by the FPGA (NI
7852R) system controlling the whole sequence. The synchronization
(waiting for mains trigger) happens during the trapping period of
each experimental repetition effectively providing no-idle-time operation.
The slow (below 50 Hz) magnetic field fluctuations are compensated
by  feedback mechanism that is enabled at the end of each sequence
repetition. This is achieved by repumping the atoms to the $m_{F}=-2$
sublevel, switching off the GEM gradient and probing the atomic spin
precession with linearly polarized probe beam at ($F=2)\to(F=1)$
transition illuminating the whole ensemble from the side. The polarization-rotation
signal is registered by differential photodiode and the precession
frequency is estimated in real time by fitting a quadratic function
around the maximum of registered signal Fourier spectrum. From this,
the frequency-error signal is obtained and fed to PI (proportional-integral) controller that
modulates the current in the $z$-direction compensation coils (see
Fig. \ref{fig:Experimental-setup}\textbf{a} for reference). With the compensation mechanism active, we
achieve long-term rms stability of 130 Hz of the Larmor frequency,
corresponding to 18.5 nT. This translates to the stability of the
relative frequency centroid of twice the Larmor frequency, i.e. 260
Hz, which is well below the desired resolution.

\subsubsection*{Signal pulse preparation}

\begin{figure*}
\includegraphics[width=1\textwidth]{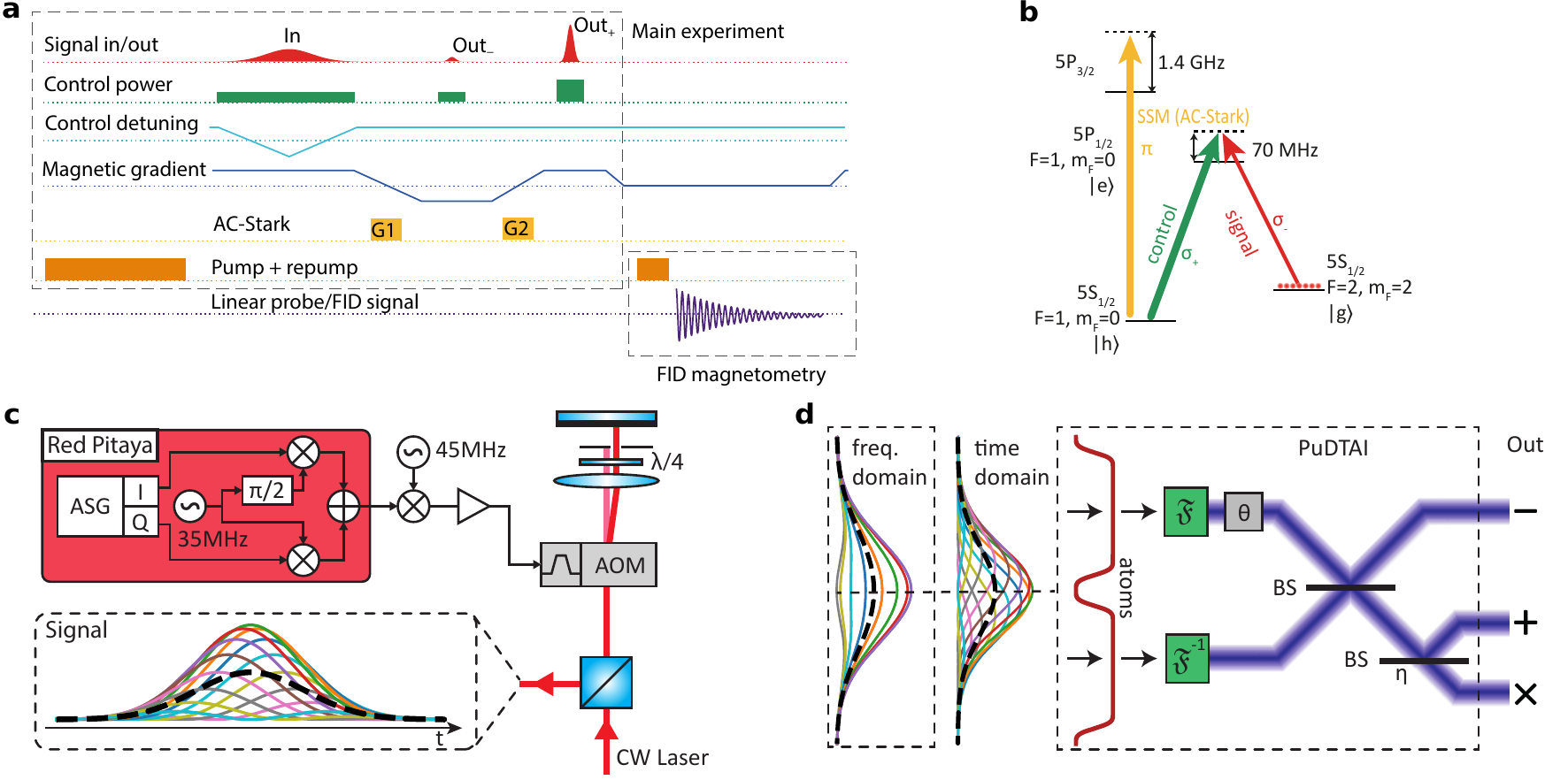}

\caption{\textbf{Experimental details and schematic representation of \sasin\ model.}
\textbf{a} The sequence is separated into the main experiment part
and the auxiliary magnetic field measurement part via free-induction
decay (FID) measurement. In the main sequence, we show the signal
input/output, the control field intensity and detuning, the GEM magnetic
field gradient, the AC-Stark patterns with two grating patterns G1
and G2, and the pump/repump lasers that prepare the initial atomic
state. In the FID magnetometry sequence we show the same preparation
as in the main experiment as well as the FID signal registered on
a polarization rotation detector. The signal is used to correct the
mean magnetic field asynchronously. \textbf{b }The essential levels
of $^{87}$Rb atoms used in the protocol as well as control, signal
and AC-Stark light fields with respect to atomic level structure. \textbf{c} Experimental setup used to generate frequency-domain double-Gaussian
signal $\mathcal{\tilde{S}}_{\varphi}(\omega)$. The in-phase (I)
and quadrature (Q) signal components generated using the arbitrary signal generator module (ASG) are mixed with intermediate carrier
frequency at 35 MHz using digital IQ-mixer. The resulting radio-frequency
(RF) signal is then up-converted by 45 MHz with help of analog mixer.
Finally, the RF signal with carrier frequency at 80 MHz is feed into
acousto-optic modulator (AOM) in double pass configuration that carves the
light pulses (Signal) from continuous wave (CW) laser. \textbf{d} The input signal
passes through temporal aperture (atoms) and is split into two halves. The
upper and lower parts are Fourier transformed ($\mathfrak{F}$) (inverse
transform - $\mathfrak{F}^{-1}$ - for the lower part) and combined
on the beamsplitter (BS) with relative phase set to $\theta$. The
first output port ($\mathrm{Out_{-}}$) is directly detected while
the second port ($\mathrm{Out_{+}}$) before detection experiences
additional loss due to lossy ($\eta$) implementation
of the beamsplitter.\label{fig:Experimental-timing-sequence}}
\end{figure*}

The signal and control pulses are carved out from two branches of
a continuous wave frequency-stabilized \citep{Lipka2017} laser. The
first branch, used for signal pulse generation is derived by frequency-shifting
by $6834$ MHz part of laser light using EOM and Fabry–Pérot filtering
cavity (see \citep{Parniak2019} for details). Then the signal pulses
are carved with double-pass AOMs feed by home-made arbitrary waveform
generator consisting of \emph{Red Pitaya} running the \emph{PyRPL}
\citep{Neuhaus2017} software and external frequency mixer providing
upconversion of the carrier frequency to the desired 80 MHz. The setup
is depicted in Fig. \ref{fig:Experimental-timing-sequence}\textbf{c}. The \emph{Red Pitaya} with \emph{PyRPL}
allows us to prepare arbitrary (up to bandwidth limitations of 50
MHz) complex envelopes that are internally multiplied by the carrier
waveform at 35 MHz using digital IQ mixer. The signal is then externally
mixed with 45 MHz local oscillator from Direct Digital Synthesizer
controlled by the main FPGA system to match the AOM central frequency
of 80 MHz and provide extended amplitude dynamic range. 

The frequency-domain double-Gaussian signal $\mathcal{\tilde{S}}_{\varphi}(\omega)$
is generated by programming a temporal envelope $\mathcal{S}_{\varphi}(t)=\mathcal{S}\cos(\frac{\delta\omega t-\varphi}{2})\exp(-t^{2}\sigma^{2})$
with given $\delta\omega=\sigma\varepsilon$. During the measurements,
to make the virtual sources mutually incoherent we continuously ($\mod2\pi)$
change the phase $\varphi$ from $-\pi$ to $\pi$ in a way that for
single measurement the number of complete $2\pi$ cycles is in order
of few thousands.

\subsubsection*{\sasin\  Model and calibrations}

 The schematic representation of the theoretical model is depicted
in Fig. \ref{fig:Experimental-timing-sequence}\textbf{d}. The model follows  image-inversion interferometer
description with two mutually incoherent weak ($\bar{n}\ll1$) sources
at the input. 

\begin{figure}
\includegraphics[width=1\columnwidth]{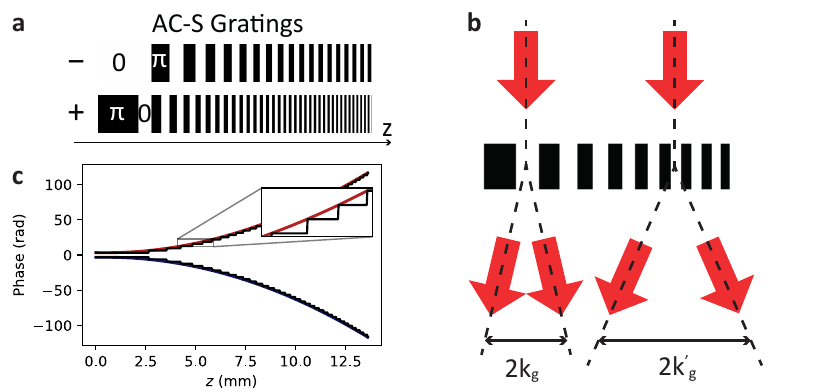}\caption{\textbf{Chirped square-wave phase grating used to simultaneously apply positive
and negative quadratic phase in space.} \textbf{a} - Two gratings used to implement
the antisymmetric ($-$) and symmetric ($+$) \sasin\ ports. \textbf{b} The
coherence at longitudinal positions $z$ is split into diffraction
orders that are separated by the grating local wavevector $k_{g}=\kappa z$.
\textbf{c} Interpretation of the chirped square-wave grating as a superposition
of positive and negative quadratic Fresnel phase profiles.\label{fig:chirped_grating}}
\end{figure}

By design, our interferometer superimposes the two arms in the Fourier
space. This can be most easily understood by investigating the Wigner
function transformations of the two signal pulse halves during propagation
through the interferometer. Let us focus on the first half that has
a temporal amplitude: 
\begin{equation}
\mathcal{A}_{-}(t)=\begin{cases}
\mathcal{A}(t) & t<0\\
0 & t\geq0
\end{cases}.
\end{equation}
The time-dependent two photon detuning which for $t<0$ is $\delta(t)=\alpha t$
virtually applies a temporal phase profile $\mathcal{A_{-}}(t)\to\mathcal{A}_{-}(t)\exp\left(i\frac{\alpha}{2}t^{2}\right)$
that corresponds to $z\to z'=z-\alpha k_{z}$ transformation of the
associated atomic coherence Wigner function $\mathcal{W}_{-}(z,k_{z})$.
In the language of temporal imaging this operation is known as propagation
through temporal lens with the focal length $f_{t}=\omega_{0}/\alpha$
\citep{Mazelanik2020} where $\omega_{0}$ is the optical carrier
frequency. The SSM modulation with spatially chirped square grating
with wavevector $k_{g}=\kappa z$ splits the coherence into multiple
diffraction orders as depicted in Fig. \ref{fig:chirped_grating}. However, as the depth of the grating is $\pi$
the 0-th diffraction order vanishes and the coherence is mostly split
to $\pm1$ orders (ideally, only 18\% is distributed into higher orders).
In the case of the first half of the signal pulse we are interested
in the $-1$-st order, that when isolated can be understood as a result
of quadratic phase modulation in the real space: $\varrho_{hg}(z)\to\varrho_{hg}(z)\exp(i\frac{\kappa}{2}z^{2})$.
This correspond to the $k_{z}\to k_{z}'=k_{z}\pm\kappa z$ transformation
of the Wigner function $\mathcal{W}_{-}(z,k_{z})$ and in terms of
temporal imaging represents temporal propagation by a distance $d_{t}=\kappa/\omega_{0}\beta^{2}$
where $\beta$ is the Zeeman shift gradient slope that facilitates
the spectrum-to-space mapping. With $\kappa=\nicefrac{1}{\alpha}$
that means $d_{t}=f_{t}$ the full transformation reads: 
\begin{alignat}{1}
k_{z} & \to k_{z}'=\kappa z\\
z & \to z'=z-\frac{1}{\kappa}k_{z}
\end{alignat}
 which up to the additional temporal phase modulation represents a
counter-clockwise $90^{\circ}$ rotation of the phase space, that
in fact is a backward Fourier transform. Similarly for the second
half of the pulse (for $t\geq0$), as the sign of $\alpha$ changes
we get a clockwise rotation and thus forward Fourier transform. Finally,
as at the input we have the full signal pulse $\mathcal{A}(t)=\mathcal{A}_{-}(t)+\mathcal{A}_{+}(t)$
the two components interfere in the in the Fourier domain when the
positive frequency components of the first half are superimposed onto
the negative components of the second half thus implementing the inversion
interferometer. Additionally, the by design Fourier transform feature
of our \sasin\  interferometer can improve the signal-to-noise ratio
when using noisy detectors as long signal pulses result in short pulses
at the interferometer output.

\begin{figure*}
\includegraphics[width=1\textwidth]{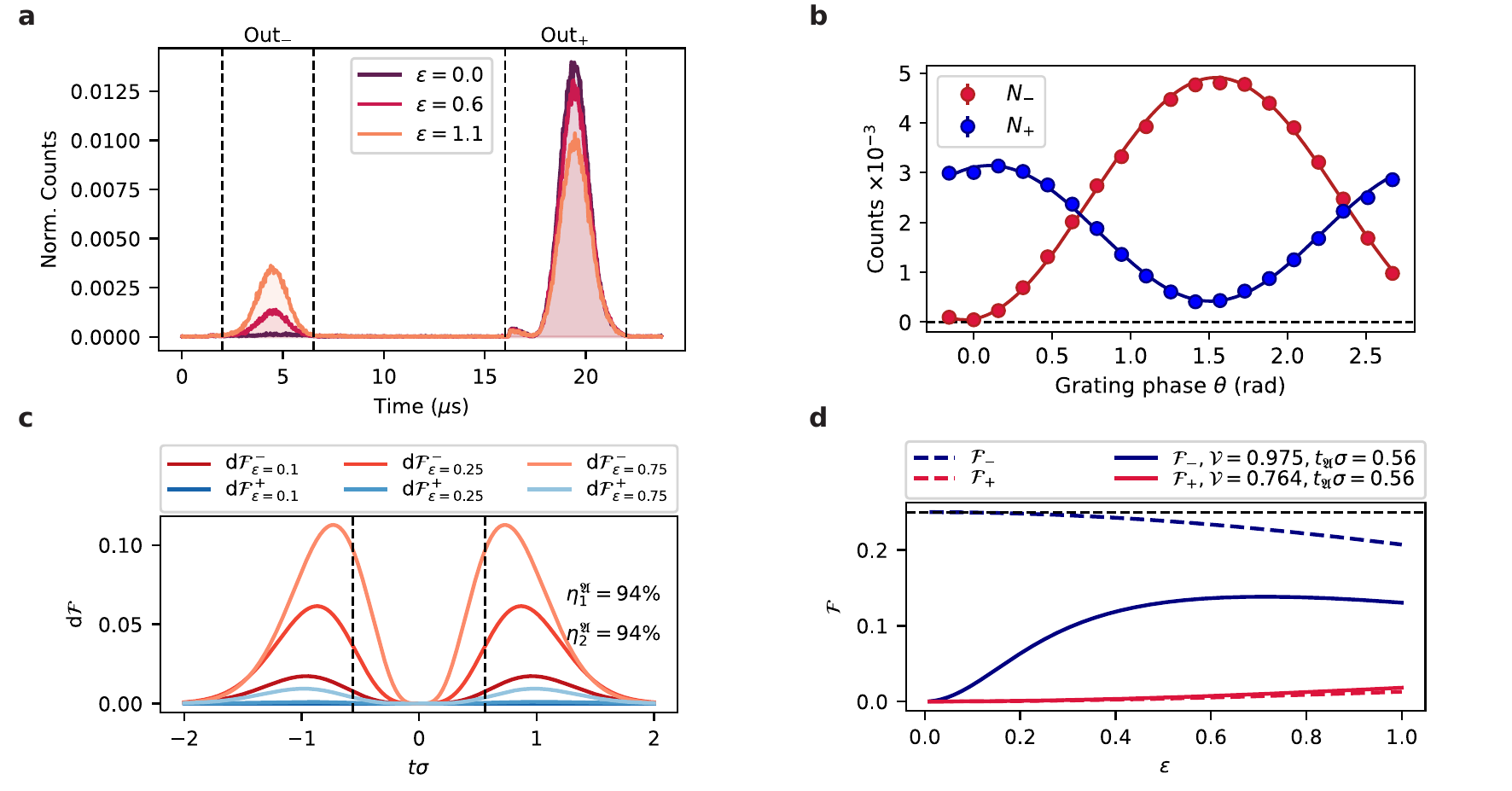}\caption{\textbf{Time-bin outputs of the \sasin\ interferometer and associated FI.} \textbf{a} Single-photon
counts distributions registered at each port ($\mathrm{Out_{-}}$, $\mathrm{Out_{+}}$) of the \sasin\  interferometer
for several chosen true separation values $\varepsilon$. \textbf{b} The
\sasin\  interferometer phase $\theta$ scan for fixed separation
parameter $\varepsilon=0$ along with fitted sine and cosine functions corresponding to two output ports ($N_{-}$ and $N_{+}$). \textbf{c} Fisher information density $\mathcal{\mathrm{d}F}_{\varepsilon}$
for direct detection of the antisymmetric ($-$) and symmetric ($+$)
subspace with the realistic parameters $\mathcal{V}_{-}$ and $\mathcal{V}_{+}$.
For small $\varepsilon$ the information is concentrated in the two
lobes outside the central part of the spectrum. The relative efficiencies
$\eta_{-}^{\mathfrak{A}}$, $\eta_{+}^{\mathfrak{A}}$ correspond to
the hard aperture (marked as dashed lines) that removes the input
signal located within the $t\sigma=\pm0.564$ range.  \textbf{d} Fisher information available at each output port of the \sasin\ device. For small separations $\varepsilon\ll1$ the antisymmetric
port ($\mathcal{F}_{-}$ corresponding to $p_{-}$) contains most
information about $\varepsilon$. The dashed lines represents the
ideal case of $\mathcal{V}_{\pm}=1$ and infinite aperture.\label{fig:porty}}
\end{figure*}

With the separation parameter $\varepsilon=\nicefrac{\delta\omega}{\sigma}$
and temporal representational of the input signal given by:
\begin{equation}
\mathcal{S}_{\varphi}(t)=\psi_{\hgz}(t)\sqrt{2}\cos(\frac{\delta\omega t-\varphi}{2})\exp(\frac{i\varphi}{2})
\end{equation}

that at the interferometer input is clipped by the (symmetric) temporal
aperture function $f_{\mathfrak{A}}(t)$, we may write the following
amplitudes for the antisymmetric and symmetric ports:

\begin{align}
u_{-}(t)= & \frac{1}{2}f_{\mathfrak{A}}(t)(\mathcal{S}_{\varphi}(t)-\mathcal{S}_{\varphi}(-t)),\label{eq:asym}\\
u_{+}(t)=\frac{1}{2} & f_{\mathfrak{A}}(t)(\mathcal{S}_{\varphi}(t)+\mathcal{S}_{\varphi}(-t)),\label{eq:sym}.
\end{align}
As the interferometer internally performs a Fourier transform of the input signal, at the
device output we observe the counts distributions given by these two
components in the frequency domain $\tilde{p}_{i}(\omega)=|\tilde{u}_{i}(\omega)|^{2}$
(see Fig. \ref{fig:porty}\textbf{a} for example of measured counts distributions,
with $\omega=\alpha t$).

The total contributions in the antisymmetric and symmetric ports
are calculated as $p_{i}=\int\tilde{p}_{i}(\omega)\mathrm{d}\omega$
which corresponds to using a spectral bucket (no frequency information)
detector. After introducing non-perfect interference visibilities
$\mathcal{V}_{-}$ and $\mathcal{V}_{+}$ and additional losses in
the symmetric port $\eta_{+}$ we arrive at outcome probabilities
that for hard Cassegrain-type aperture function given by $f_{\mathfrak{A}}(t)=\Theta(-t_{\mathfrak{A}}-t)+\Theta(-t_{\mathfrak{A}}+t)$
take the form:
\begin{align}
p_{-} & =\frac{1}{2}\left(\mathrm{\mathrm{erfc}(\sqrt{2}\sigma t_{\mathfrak{A}})}-\mathcal{V}_{-}e^{-\frac{\varepsilon^{2}}{8}}f(t_{\mathfrak{A}},\varepsilon)\right),\label{eq:prob1}\\
p_{+} & =\frac{\eta_{+}}{2}\left(\mathrm{\mathrm{erfc}(\sqrt{2}\sigma t_{\mathfrak{A}})}+\mathcal{V}_{+}e^{-\frac{\varepsilon^{2}}{8}}f(t_{\mathfrak{A}},\varepsilon)\right),\label{eq:prob2}\\
p_{\times} & =1-p_{-}-p_{+},\label{eq:prob3}
\end{align}
where $f(t_{\mathfrak{A}},\varepsilon)=\frac{1}{2}\left(\mathrm{erfc}\left(\frac{4t_{\mathfrak{A}}\sigma-i\varepsilon}{2\sqrt{2}}\right)+\mathrm{erfc}\left(\frac{4t_{\mathfrak{A}}\sigma+i\varepsilon}{2\sqrt{2}}\right)\right)$. The
$p_{-},\,p_{+}$ are probabilities of detecting photon in the antisymmetric
and symmetric interferometer port respectively, while $p_{\times}$
is a probability of no detection event. In Fig. \ref{fig:porty}\textbf{b} we show the contributions to the symmetric and antisymmetric port for $\varepsilon=0$ when varying the grating phase $\theta$, which demonstrates the high interference visibility.

The atomic cloud size along with the strength of the magnetic field
gradient and slope of the control field chirp $\alpha$ limits the
temporal aperture $\mathfrak{A}$ of the interferometer. We design
the aperture to contain most informationally valuable parts of input
signal while at the same reducing the impact of the instantaneous
control chirp reversal at $t=0$. This reversal by being inherently
broadband spoils the time-to-position mapping by virtually locally
broadening the input signal. To get rid of this effect we chose the
central two photon detuning $\delta(t=0)$ to be outside the magnetically
broadened absorption spectrum providing no light-to-atoms mapping
at the $\alpha$ reversal time. This forms an Cassegrain type aperture
described by $f_{\mathfrak{A}}$ that limits mostly the efficiency
in the second interferometer port, that for $\varepsilon\ll1$ contains
no information, and serves only as brightness reference for the first
port. We show this in Fig. \ref{fig:porty}\textbf{c}, where we plot the
Fisher information densities  defined as: 
\begin{equation}
\mathrm{d}\mathcal{F}_{i}=\frac{1}{p_{i}(t)}\left(\frac{\partial}{\partial\varepsilon}p_{i}(t)\right)^{2}\mathrm{d}t.\label{eq:fidens}
\end{equation}

The impact of the aperture on photon flux efficiencies in the symmetric
and antisymmetric port can be investigated by comparing the photon
detection probabilities (Eq. \ref{eq:prob1} and Eq. \ref{eq:prob2})
with the case of infinite aperture $t_{\mathfrak{A}}=0$ and taking
the limit of $\varepsilon\to0$. This gives $\eta_{p_{-}}^{\mathfrak{A}}\approx\mathrm{erfc}(\sqrt{2}\sigma t_{\mathfrak{A}})+2\sqrt{\nicefrac{2}{\pi}}\sigma t_{\mathfrak{A}}\exp(-2\sigma^{2}t_{\mathfrak{A}}^{2})$
and $\eta_{p_{+}}^{\mathfrak{A}}\approx\mathrm{erfc}(\sqrt{2}\sigma t_{\mathfrak{A}})$
which for realistic aperture $t_{\mathfrak{A}}=0.564/\sigma$ gives
$\eta_{p_{-}}^{\mathfrak{A}}\approx0.74$, $\eta_{p_{+}}^{\mathfrak{A}}\approx0.26$.
However, in the context of $\varepsilon$ estimation those are not
crucial, and we should look how the FI changes with the $t_{\mathfrak{A}}$.
The FI density \ref{eq:fidens} integrated over the aperture gives
the total FI available at the outputs of the device $\mathcal{F}_{i}^{\mathfrak{A}}=\int_{\mathfrak{A}}\mathrm{d}\mathcal{F}_{i}$
that can be utilized when using a spectrally-resolved detection. Notably,
in the case of infinite aperture and ideal visibility the total FI
at the output $\mathcal{F}_{\Sigma}^{\infty}=\mathcal{F}_{-}^{\infty}+\mathcal{F}_{+}^{\infty}$
is independent of $\varepsilon$ and equals QFI which means that shapes
of the distributions at each port of the device contain additional
information that can improve the resolution for larger $\varepsilon$
values as has been previously recognized in the context of real-space
imaging \citep{Nair2016a}. To investigate the impact of the finite
aperture $\mathfrak{A}$ we calculate the FI $\mathcal{F}_{\pm}^{\mathfrak{A}}$
and evaluate the efficiencies as $\eta_{\mathcal{F_{\pm}}}^{\mathfrak{A}}=\mathcal{F}_{\pm}^{\mathfrak{A}}/\mathcal{F}_{\pm}^{\infty}$
which in the case of our aperture and visibilities both approximately
equals 94\% as included in Fig. \ref{fig:porty}\textbf{c}. In the case of spectral
bucket detector, the total Fisher information calculated for the observable
outcome probabilities (\eqref{eq:asym},\eqref{eq:sym}) takes the
form:
\begin{gather}
\label{eq:fisher-pudtai}
\mathcal{F}_{\mathrm{PuDTAI}}=\underbrace{\frac{\mathcal{V}_{-}^{2}F}{64p_{-}}}_{\mathcal{F}_{-}}+\underbrace{\frac{\eta_{+}^{2}\mathcal{V}_{+}^{2}F}{64p_{+}}}_{\mathcal{F}_{+}},\\
F=e^{-\varepsilon^{2}/4}\left(\sqrt{8/\pi}e^{-2t_{\mathfrak{A}}^{2}\sigma^{2}+\varepsilon^{2}/8}\sin(t_{\mathfrak{A}}\sigma\varepsilon)+\varepsilon f(t_{\mathfrak{A}},\varepsilon)\right)^2.
\end{gather}
In Fig. \ref{fig:porty}\textbf{d} we plot the two parts of $\mathcal{F}_{\mathrm{PuDTAI}}$ for both the ideal and realistic case.

Interestingly, when we now evaluate the FI efficiencies in the similar
way as in the case of spectrally-resolved detection and with the finite
visibilities we may observe improvement over the apertureless case
for certain $\varepsilon$ range. This is direct result of filtering
the informationally not valuable central part of the signal that contains
most of the leaking (due to $\mathcal{V}<1$) photons. The improvement
is most prominent for $\varepsilon\to0$ and in our case reads $\eta_{-}^{\mathfrak{A}}=\eta_{+}^{\mathfrak{A}}\approx2.1$.
The experimental parameters $\mathcal{V}_{-}\approx(97.51\pm0.03)\%$, 
$\mathcal{V}_{2}=(76.4\pm0.8)\%$, $t_{\mathfrak{A}}\sigma\approx0.564\pm0.002$
and the efficiency $\eta_{+}\approx0.719\pm0.009$ are obtained from
calibration measurements consisting on running the protocol for different
$\varepsilon$ and interferometer phase $\theta$ as shown in Fig. \ref{fig:porty}\textbf{b}. 

Finally, we can also evaluate the mean efficiencies of storage retrieval
from the memory. We obtain $\eta_{p_{-}}=0.83\%$ and $\eta_{p_{+}}=0.60\%$.
We note that those efficiencies are very similar to the ones obtained
in the QPG protocol \citep{Donohue2018}. To boost the efficiencies
of the GEM one could design a larger and more dense atomic ensemble
or prepare the ensemble inside a cavity resonant with the signal field
that effectively boosts the optical depth and thus the efficiency
\citep{PhysRevLett.124.210504,PhysRevLett.126.090501,doi:10.1063/1.5065431}.

\subsubsection*{Maximum likelihood estimation}

The estimation of the separation parameter $\varepsilon$ in both cases (QMTI and PuDATI) follows the standard maximum likelihood estimation procedure. In the case of QMTI the single outcome probability function is given by 
\begin{equation}
p_{\varepsilon}(\omega)=\tilde{I}(\omega)=\frac{1}{2}\left(|\tilde{\psi}(\omega-\sigma\varepsilon/2)|^{2}+|\tilde{\psi}_{-}(\omega+\sigma\varepsilon/2)|^{2}\right). 
\end{equation} 
For given distribution of measurement outcomes $\boldsymbol{\omega}=(\omega_{1},...,\omega_{n})$ we construct the likelihood function 
\begin{equation}
\mathcal{L}_{\boldsymbol{\omega}}(\varepsilon)=\prod_{i=1}^{n}p_{\varepsilon}(\omega_{i}),
\end{equation}
that is numerically maximized to yield the maximum likelihood value of $\varepsilon$. For \sasin\ we have the three possible outcomes that are characterized by the three probabilities: $p_{-},p\,_{+},\,p_{\times}$. For N processed photons we have the likelihood given by a probability mass function of the trinomial distribution: 
\begin{multline}
\mathcal{L}_{N_{-},N_{+}}(\varepsilon)=\frac{N!}{N_{-}!N_{+}!(N-N_{-}-N_{+})!}\times\\
p_{-}(\varepsilon)^{N_{-}}p_{+}(\varepsilon)^{N_{+}}p_{\times}(\varepsilon)^{N-N_{-}-N_{+}},
\end{multline}
which is maximized for $p_{-}(\varepsilon)/p_{+}(\varepsilon)=N_{-}/N_{+}$. From this, we estimate the separation by numerically solving the formula for $\varepsilon$.
\subsubsection*{Superresolution parameter}

As given by Eq. \ref{eq:limitS} the superresolution parameter $\mathfrak{s}$ must be evaluated at a vanishing separation $\varepsilon$.  For the \sasin\ protocol we directly plug in the formula for Fisher information (Eq. \ref{eq:fisher-pudtai}) to the limit (Eq. \ref{eq:limitS}) and obtain:
\begin{multline}
\mathfrak{s}_{\mathrm{\sasin}} =\\ \frac{e^{-4 \sigma ^2 t_\mathfrak{A}^2}  \left(\mathcal{V}_-^2 (\mathcal{V}_++1)-\eta_+ \mathcal{V}_- \mathcal{V}_+^2+\eta_+ \mathcal{V}_+^2\right)}{4 (\mathcal{V}_--1) (\mathcal{V}_++1) \left(\text{erf}\left(\sqrt{2} \sigma  t_\mathfrak{A}\right)-1\right)} \times \\  \left(-e^{2 \sigma ^2 t_\mathfrak{A}^2} \text{erf}\left(\sqrt{2} \sigma  t_\mathfrak{A}\right)+e^{2 \sigma ^2 t_\mathfrak{A}^2}+2 \sqrt{\frac{2}{\pi }} \sigma  t_\mathfrak{A}\right)^2
\label{eq:sasinS}
\end{multline}
The value may be evaluated accurately with an uncertainty, as we know all the relevant experimental parameters. Finally, if we set $\mathcal{V}_- = \mathcal{V}_+ =\mathcal{V}$ and $\eta_+ = 1$ and $t_\mathfrak{A} = 0$, we obtain $\mathfrak{s}_\mathrm{\sasin} = \mathfrak{s}_\mathrm{SLIVER}$ (see Eq. \ref{eq:sliverV}).

The classical (DI) spectrometers such as Grating, FT and QMTI that
are not tailored to the given signal mode function $\tilde{\psi}(\omega)$
can be directly used with signals that have different FWHM or $\sigma$.
To indicate that, we represent them as lines covering some $\sigma$
range that depend on given spectrometer implementation. The super-resolution
parameter $\mathfrak{s}$ in this case is calculated using the same formula
as in the case of spectrometers tailored for the given $\tilde{\psi}(\omega)$
such as PuDTAI or QPG. Any DI spectrometer has its bandwidth (BWL)
and resolution limit (RL). These are causing the $\mathfrak{s}$ parameter
to drop below $1$ for signals with bandwidth exceeding the BWL or
that are narrower than RL. The limitation from the finite RL broadens
the signal mode function $\tilde{\psi}(\omega)$ and thus spoils the
separation estimation sensitivity by virtually making the $\varepsilon$
smaller. For a RL given by $\sigma_{\mathrm{RL}}$ the broadening
effect can be calculated as $\sigma\to\sqrt{\sigma^{2}+\sigma_{\mathrm{RL}}^{2}}$
which result in the $\varepsilon$ smaller by $\frac{\sigma}{\sqrt{\sigma^{2}+\sigma_{\mathrm{Rl}}^{2}}}$
factor. On the other hand the BWL characterized by $\Sigma_{\mathrm{BWL}}$
cuts the informationally valuable tails of signals with $\sigma>\Sigma_{\mathrm{BWL}}$
that results in drop of the Fisher information given as $\mathcal{F}_{\mathrm{DI}}=\int_{\Sigma_{\mathrm{BWL}}}\mathrm{d}\mathcal{F_{\mathrm{DI}}}$
with the $\mathrm{d}\mathcal{F}_{\mathrm{DI}}$ being the DI Fisher
information density and the integration goes through the whole available
bandwidth. In Fig. \ref{fig:Comparison} we plot several $\mathfrak{s}$
curves for various DI spectrometers. These are: Grating - a grating
spectrometer with gratings of lengths ranging from $1\:\mathrm{cm}$
to $10\,\mathrm{cm}$, with grating period of $1/1200\,\mathrm{mm^{-1}}$
and $\Sigma_{\mathrm{BWL}}=10^{3}\times\sigma_{\mathrm{RL}}$; FT
- Bruker IFS 125HR Fourier transform spectrometer with $0.001\:\mathrm{cm^{-1}}$
resolution and $50\times10^{3}\:\mathrm{cm^{-1}}$ and the MIR spectral
range; QMTI - the Quantum Memory Temporal Imaging Spectrometer with
$\sigma_{\mathrm{RL}}=7.2\:\mathrm{kHz}$ and $300\,\mathrm{kHz}$
BWL.

\section*{Data availability}

Data presented in Fig. 5 and Fig. 9 have been deposited in RepOD Repository for Open Data at \url{https://doi.org/10.18150/JSUKE9}. Any other data that support the findings of this study are available from the corresponding author upon reasonable request.

\section*{Code availability}

Code used in data analysis is available from the corresponding author upon reasonable request.

\section*{Acknowledgments}

We thank W. Wasilewski, M.~Jachura and R.~Demkowicz-Dobrzański for insightful discussions and K. Banaszek for the generous support.
This work has been funded by National Science Centre (Poland) grant
no. 2017/25/N/ST2/00713, Polish science budget funds for years 2017-2021 as a research project within the
``Diamentowy Grant'' program of the Ministry of Science and Education (DI2016 014846), Office of Naval
Research (USA) grant no. N62909-19-1-2127 and by the MAB/2018/4 “Quantum Optical Technologies” project.
The Quantum Optical Technologies project is carried out within the
International Research Agendas program of the Foundation for Polish
Science co-financed by the European Union under the European Regional
Development Fund. Michał Parniak and Mateusz Mazelanik were also supported by the Foundation
for Polish Science via the START scholarship.

\section*{Author contributions}

M.M. and M.P. conceived the scheme and planned the experiment and
the theoretical research. M.M., A.L. and M.P. contributed to the experimental
setup, software and data collection. M.M. analyzed the data, developed
theory and prepared figures. M.M. and M.P. wrote the manuscript. All
authors discussed the results.

\section*{Competing interests}

The authors declare no competing interests.

\end{document}